# Reversible Gas Sensing by Ferroelectric Switch and 2D Molecule Multiferroics in In$_2$Se$_3$ Monolayer

*Xiao Tang, Jing Shang, Yuantong Gu, Aijun Du, and Liangzhi Kou*[*]

School of Chemistry, Physics and Mechanical Engineering, Queensland University of Technology, Gardens Point Campus, Brisbane, QLD 4001, Australia

ABSTRACT: Two-dimensional ferroelectrics are important quantum materials which have found novel application in nonvolatile memories, however, the effects of reversible polarization on chemical reactions and interaction with environments are rarely studied despite of its importance. Here, based on the first-principles calculations, we found distinct gas adsorption behaviors on the surfaces of ferroelectric In$_2$Se$_3$ layer and the reversible gas caption and release controlled by ferroelectric switch. We rationalize the novel phenomena to the synergistic effect of the different electrostatic potential and electron transfer induced by band alignments between frontier molecular orbitals of gas and band-edge states of substrate. Excitingly, the adsorption of paramagnetic gas molecules such as NO and NO$_2$ can induce surface magnetism, which is also sensitive to ferroelectric polarization direction of In$_2$Se$_3$, indicating the application of In$_2$Se$_3$ as threshold magnetic sensors/switch. Furthermore, it's suggested two NO molecules prefer to ferromagnetically couple with each other, the Curie temperature is polarization dependent which can reach up to 50K, leading to the long-sought 2D molecule multiferroics. The ferroelectric controllable adsorption behavior and molecule multiferroic feature will find extensive application in gas caption, selective catalytic reduction and spintronic device.

Two-dimensional (2D) ferroelectrics, referring to the materials with reversible polarization under external stimuli, has been a fast-developing field which attracts intensive investigations due to their underlying new science and potential applications.[1, 2] As an extensively studied research topic on traditional $ABO_3$ bulk perovskite in the past decades,[3, 4] it is recently resurrected due to the successful fabrications/synthesis of ferroelectric (FE) materials with atomic thickness, such as $SnTe$,[5] $In_2Se_3$[6] and $CuInP_2S_6$[7] layers with in-plane or out-plane ferroelectricity, which provided the excellent platforms to explore new physics. As a result of intrinsic symmetry breaking, the Rashba effects, controllable spin vortex and FE quantum Hall phase were observed.[8, 9] The unique reversible spontaneous electric polarization and cooperative phenomena like the coexistence of ferroelectricity and ferromagnetism down to the 2D limit opens avenues for numerous novel applications of non-volatile memory nanodevices and transistors.[10-13] Taking ferroelectric random access memories (Fe-RAMs) for instance, the bistable and reversible phases can equivalently represent logic "0" and "1" states, endowing ultrafast logic operation and low-power consumption.[14] In contrast to the extensive investigations on the physics and potential application on nanodevices, only a few studies focus on the effects of polarization on chemical reactions despite of its importance. For example, recent studies have shown that 2D FE materials can find promising application in the photocatalytic reactions, where the polarization can facilitate the overall water splitting in $M_2X_3$ (M = Al, Ga, In; X = S, Se, Te).[15, 16] Janus MoSSe has been predicted as a potential wide solar-spectrum water-splitting photocatalyst with a low carrier recombination rate, and enhance/weaken the gas sensitivity due to the polarization induced by structural symmetry breaking and the controlled electron transfer.[17] However, the effects of reversible polarization in 2D FE materials on gas adsorption (capture and release) are rarely investigated despite of the highly potential.

Gas sensors and capturers with controllable adsorption manners, which can detect and remove the harmful gases from air, are regarded as the advanced technologies to solve the worldwide air pollution and greenhouse effects.[18] Due to large surface ratio and active chemical activities, 2D layered materials have been identified as the promising candidates, however, the controllability and gas desorption are still the open questions to be addressed since the tightly adsorbed gas molecule will lead to the "poisoning" of the sensing materials. Although numerous works have shown that with the aid of external fields like charge doping,[19] strain,[20] and electric field,[21] the adsorption and desorption of gas molecules can be controlled, it is still challenging to achieve precise control on the nanoscale. Due to the intrinsic reversible polarization, the 2D FE materials are the possible ideal platforms to control gas adsorption behaviors and achieve gas capture/release. The strategy, if achieved, will not only provide a feasible solution for the environmental issues, but also lead to the possible controllable catalysis like selective catalytic reduction (SCR)[22, 23] where the concentration of $NO_x$ can be controlled by the polarized surfaces.

In present work, we found the distinct gas adsorption behaviors of $NH_3$, $NO$ and $NO_2$ on FE surfaces of monolayer $In_2Se_3$, where the adsorption strength, electron transfer and electronic variation strongly depend on the polarization directions. The phenomena are well rationalized by synergistic effects of relative electrostatic potentials between substrate and molecules, and the arrangements of band-edge states and frontier molecular orbitals. By FE switch, the adsorption strength of $NH_3$ can thus be modulated from chemi- or phys-adsorption to achieve the gas adsorption and release after overcoming a small energy barrier, leading to the reversible gas sensing. For magnetic molecule $NO$ and $NO_2$ adsorption, it is found that the magnetic moment can be also controlled by the FE switch. More interestingly, the neighboring $NO$ molecules prefer to ferromagnetically couple with each other, leading to the coexistence of ferroelectricity and ferromagnetism, namely novel 2D molecule multiferroics. The Curie temperatures of the ferromagnetism on two FE sides of $In_2Se_3$ layer are significantly different, but both are up to 50K from the Monte Carlo simulation, indicating the strong magnetoelectric coupling. The newly proposed mechanism of gas sensing based on the band arrangements will deepen the understanding of the interaction between gas and substrate, which can guide the design for next generation of selective and sensitive sensors. Together with reversible gas capture/release and molecule multiferroics, these findings will open an avenue for FE sensing and spintronic devices.

**Results and Discussion.** *Reversible sensing.* We choose the FE $In_2Se_3$ monolayer as the substrate to study the adsorption behaviors of gas $NH_3$, NO and $NO_2$, while other gases including $CO_2$, CO and $SO_2$ were also investigated and discussed in present work as a supplementary. Firstly, the structural characters of monolayer $In_2Se_3$ was studied as a benchmark. The stable 2D $In_2Se_3$ monolayer exhibits quintuple layer structure stacking in the sequence of Se-In-Se-In-Se as illustrated in Figure 1a. The calculated lattice parameter is 4.06 Å and the height of $In_2Se_3$ monolayer is 6.82 Å, agreeing well with previous studies.[15, 24, 25] Due to the inversion symmetry breaking, an intrinsic electric polarization which can be switched via laterally moving the central Se layer through accessible kinetic pathways by extrinsic electric field has emerged. The calculated electrostatic potential difference of two respective surfaces is found to be 1.20 eV, confirming the existence spontaneous electric polarization of the two surfaces.

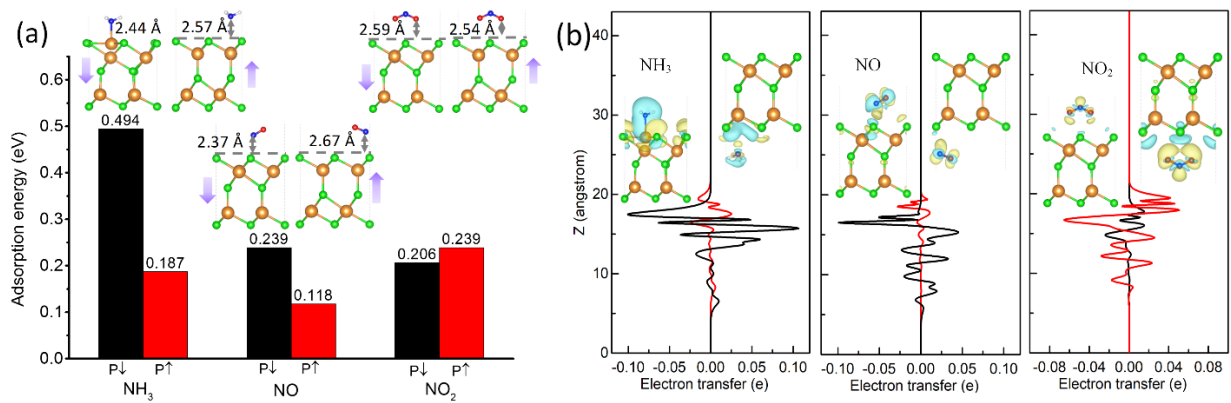

**Figure 1.** (a) Adsorption energies and (b) electron transfer of $NH_3$, NO and $NO_2$ on $In_2Se_3$ monolayer. The black and red columns (lines) are the adsorption energies (electron transfer) on the $In_2Se_3$ (P ↓) and (P ↑) surfaces, respectively. The most stable adsorption configurations together with the adsorption distance are also presented as the insets in panel a. The charge density differences after the gas adsorptions are inserted into panel b, where the cyan (yellow) indicates electron loss (accumulation) and the isosurface value is 0.0004 e Å$^{-3}$. The purple arrows indicate the directions of electric polarization. Green, orange, white, blue and red balls present the Se, In, H, N, and O atoms, respectively.

We then investigate the adsorption behaviors of $NH_3$, NO and $NO_2$ molecules on $In_2Se_3$ surfaces with different electric polarization directions pointing downwards and upwards (P ↓ and P ↑). A 2×2 supercell of $In_2Se_3$ monolayer, with a single gas molecule adsorbed to it. Several possible adsorption sites (hollow, on top of In or Se atom) and different molecular orientations (vertical, horizontal or parallel orientation) are examined to figure out the most stable adsorption position and orientation (see insets in Figure 1a).

For $NH_3$, we have identified the most stable configurations onto P ↓ and P ↑ sides of $In_2Se_3$ monolayer as shown in Figure 1. For both sides of monolayer $In_2Se_3$, the top position of In site is always the most favorably preferred while the N atoms are pointing towards the surface. However, the obviously different adsorption behaviors can be seen on the different surface. On the P ↓ side, the $In_2Se_3$ structure is distorted and the N atom of $NH_3$ is bonded with In atom in the second layer (bond length of N-In is 2.44 Å), the calculated adsorption energy is -0.494 eV, which implies a chemical adsorption. In contrast, at the opposite surface of $In_2Se_3$ (P ↑), geometry of $In_2Se_3$ is not severely distorted after $NH_3$ adsorption and there is no chemical bonding between the substrate and the molecule, the adsorption energy is relatively moderate (-0.187 eV) which is in the range of physical adsorption. Calculated charge density difference also confirms that in the case of P ↓ side, more charges have been transferred between $NH_3$ and $In_2Se_3$ (see Figure 1b).

Upon exposure NO to the $In_2Se_3$ sheet (Figure 1a), the results obtained suggest that the NO adsorbs via N atom on both sides. For NO adsorption on $In_2Se_3$ (P ↓), the $E_{ads}$ is -0.190 eV and the distance between N atom to the upper Se surface is 2.37 Å. The bond length of NO is slightly shortened to 1.165 Å compared to the free NO. According to the Bader charge population, in this situation ~0.04e is transferred from NO molecule to $In_2Se_3$. When NO adsorbed on $In_2Se_3$ (P ↑), the $E_{ads}$ is -0.118 eV with a charge transfer of ~0.02e from $In_2Se_3$ to NO, leading to a weak interaction. The different adsorption performance can be also seen from the charge density difference indicated in Figure 1b (insets in middle panel). In the case of $NO_2$ adsorption, the $E_{ads}$ is -0.205 eV on $In_2Se_3$ (P ↓) but -0.239 eV on $In_2Se_3$ (P ↑). Consistent with the adsorption energy difference, Bader analysis also indicate that, more charge (~0.15 e) is transferred from $In_2Se_3$ (P ↑) to $NO_2$ molecule compared with that (~0.07 e) on $In_2Se_3$ (P ↓). Therefore, it is clear that $NH_3$ and NO are preferring to adsorb on $In_2Se_3$ (P ↓), whereas $NO_2$ molecule are more likely to be adsorbed on $In_2Se_3$ (P ↑), indicating that the adsorption behaviors on the surface of $In_2Se_3$ monolayer can be well regulated with FE switch.

To check the dynamics and temperature effects of adsorption behaviors on two FE surfaces, we estimated the recovery time (τ) based on the transition state theory and van't-Hoff-Arrhenius expression[26] which are given by:

$$\tau = v_0^{-1} \exp(\frac{-E_{ads}}{k_B T})$$

Where $v_0$ is the attempt frequency ($10^{12}$ s$^{-1}$ for $NH_3$, NO and $NO_2$)[27-29], T is temperature and $k_B$ the Boltzmann's constant ($8.62 \times 10^{-5}$ eV/K), $E_{ads}$ is the adsorption energy on the different surface as calculated above. At room temperature (298 K), it is found that the recovery time of $NH_3$ is 5 magnitude order faster on P ↑ than that of P ↓ surface ($1.45 \times 10^{-9}$ vs $2.25 \times 10^{-4}$ s). The recovery times for NO and $NO_2$ on two ferroelectric surfaces are relatively close, but there are still 1-2 magnitude order differences ($9.89 \times 10^{-11}$ vs $1.63 \times 10^{-9}$ s for NO; $1.10 \times 10^{-8}$ vs $2.92 \times 10^{-9}$ s for $NO_2$). The obvious differences of recovery times on the opposite surfaces indicate the possibility of gas capture and release controlled with the environmental fluctuations and temperature.

*Underlying mechanism for the distinct adsorption behavior*. The distinct adsorption behaviors (electron transfer and adsorption strength) at two surfaces of FE $In_2Se_3$ monolayer can be rationalized with the synergistic effect of the electrostatic potential difference and the relative band alignment between gas molecule and the $In_2Se_3$ monolayer. Because of the asymmetric structure, the electrostatic potential on the two surfaces of $In_2Se_3$ monolayer are different, which are 4.63 and 5.83 eV respectively relative to the Fermi level (Figure S1). For the adsorbed molecules, the adsorption energies are positively correlated with the electrostatic potential differences between adsorbents and substrate. Based on the adsorption orientation, we compared the potential value of molecule with those of $In_2Se_3$ monolayer as shown in Figure 2a. It is found that the potential differences for $NH_3$ are 0.97 eV and 0.23 eV respectively when it is absorbed on the P ↓ and P ↑ surface. The corresponding values are 1.22 and 0.02 eV for NO, 0.01 and 1.21 eV for $NO_2$. Larger potential difference between molecule and substrate is, more strongly the gas molecule can be absorbed. It is therefore easy to understand the adsorption difference on the FE surface of $In_2Se_3$ monolayer, and the adsorption preference ($NH_3$ and NO on P ↓ while $NO_2$ on P ↑ ).

Besides the potential difference, the deeper mechanism can be attributed to the different electron transfer originated from the band alignment. Similar to the studies of photocatalysis of water splitting and photovoltaic solar cells, we shift the vacuum level on both sides to zero ($E_{vac}$=0 eV).[30] As shown in Figure 2b, the band levels of $In_2Se_3$ (P ↓ ) lie about 1.20 eV (work functional as calculated by PBE) below the corresponding band levels of $In_2Se_3$ (P ↑ ) due to the intrinsic electric field. Therefore, when gas molecules are absorbed on $In_2Se_3$ [$In_2Se_3$ (P ↓ ) or $In_2Se_3$ (P ↑ )], different band alignments will endow the tunability of electron transfer. As an evidence, a recent study shows that, redox potentials of $H_2O/O_2$ and $H^+/H_2$ between the two surfaces of $III_2V_3$ monolayer are shifted as a result of vacuum level difference, benefitting to photocatalytic water splitting.[15] Hence, it is reasonably expected that as gas molecules are exposed to FE $In_2Se_3$, the band arrangements between the highest occupied molecular orbitals (HOMO)/lowest unoccupied molecular orbitals (LUMO) of gas and conduct band minimum (CBM) / valence band maximum (VBM) of $In_2Se_3$ would be different, which offers new avenues for realizing gas adsorption/release and controllable catalysis like SCR.

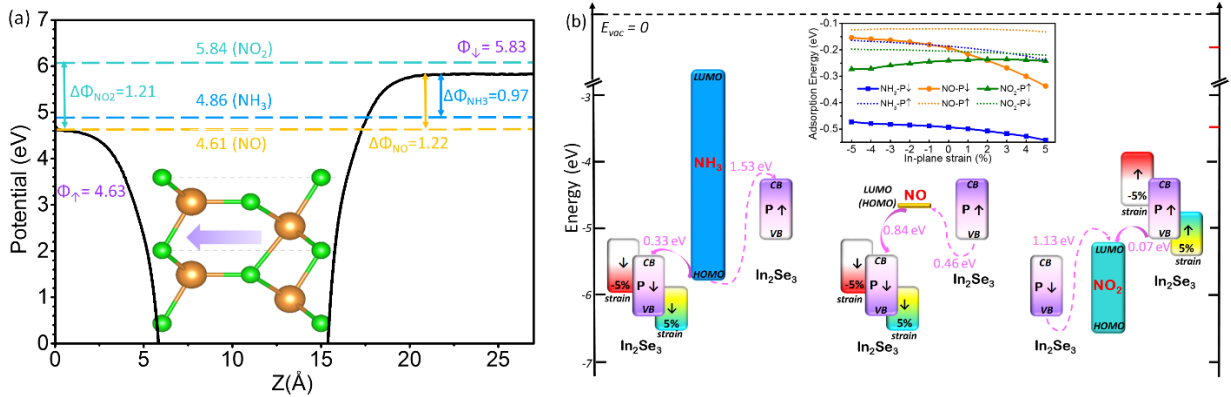

**Figure 2.** (a) Electrostatic potential of In$_2$Se$_3$ surfaces and the potential differences with adsorbed gas molecules; (b) Band alignments between the frontier molecular orbitals of gases and band edge states of In$_2$Se$_3$ with different polarization direction, all the energy levels are shifted relative to the Fermi level. Band alignments of In$_2$Se$_3$ under in-plane strain (±5%) are also shown for comparison. Inset in panel (b) is the adsorption energies of NH$_3$, NO and NO$_2$ adsorbed on In$_2$Se$_3$ (P ↓) and In$_2$Se$_3$ (P ↑), as a function of applied in-plane strain. Solid lines highlight the adsorption preference (NH$_3$ and NO on P ↓ while NO$_2$ on P ↑). Dash lines denote the case of gas adsorption on the opposite surface.

When the NH$_3$ molecule is exposed to In$_2$Se$_3$, as illustrated in Figure 2b, the HOMO of NH$_3$ lies closer to the CBM edge state of In$_2$Se$_3$ (P ↓), which indicates that electron transfer from molecule to the substrate should occur more easily in this scenario. Therefore, it is likely that NH$_3$ molecule is preferred to adsorb on the In$_2$Se$_3$ (P ↓) rather than (P ↑) surface with stronger adsorption strength. Upon adsorption of NO molecule, CBM state of In$_2$Se$_3$ is 0.84 eV lower than the HOMO of NO when it is adsorbed on the surface of In$_2$Se$_3$ (P ↓), leading to direct Z-scheme charge transfer. However, when NO is adsorbed on the opposite surface, the HOMO (LUMO) is located above (below) VBM (CBM) of In$_2$Se$_3$, leading to type I band alignment but possess weaker interaction between adsorbent and substrate due to smaller electron transfer.[31] Therefore, NO also prefers to be adsorbed on the top surface of $In_2Se_3$ (P ↓), but not (P ↑). It is worth noting that the HOMO of NO is degenerate and is half-filled, therefore it is also the LUMO, which is in agreement with previous studies.[32] Similar mechanism can be applied to $NO_2$ when it is adsorbed on $In_2Se_3$ (P ↑) since LUMO states for $NO_2$ are lower than the VBM of $In_2Se_3$ (P ↑). In the case of NO and $NO_2$ adsorption, direct electron transfer as a result of Z-scheme band alignment is favorable. The results reflect therefore, $NH_3$ and NO are preferred to be adsorbed on $In_2Se_3$ (P ↓) whereas $NO_2$ tends to $In_2Se_3$ (P ↑) surface. These analysis from the proposed band alignments matches very well with the calculated adsorption energy and electron transfer for NO, $NO_2$ and $NH_3$ on $In_2Se_3$. The distinct adsorption behaviors on the different surface were also recently observed in Janus MoSSe,[33] which can contribute to the different band alignments at two opposite surfaces due to the electrostatic potential difference (vacuum level). The validness of the newly proposed mechanism can be further verified by the adsorption behavior of CO, $CO_2$ and $SO_2$ (Figure S2). It is found that the energy level HOMO (LUMO) of both CO and $CO_2$ are much lower (higher) than the VBM (CBM) of $In_2Se_3$, therefore forming the straddling band alignment relative to band edge states.[31] The large difference between the band-edge state and frontier molecular orbitals renders the trivial electron transfer, weak interactions and insignificant adsorption strength difference on two FE surfaces. For toxic $SO_2$, it is still a straddling band alignment relative to $In_2Se_3$ regardless of P ↓ or P ↑. However, HOMO of $SO_2$ is obviously closer to CBM state of $In_2Se_3$ with P ↓, therefore $SO_2$ is prone to be adsorbed on the surface with polarization pointing downwards, the expectation is well consistent with the adsorption behavior calculations, see Figure S2.

It is well known that the adsorption strength and electron transfer of gases on 2D materials can be adjusted by in-plane strain, which can improve the gas sensitivity of these candidates.[33] The proposed mechanism of band alignments can be also used to give a reasonable explanation for the phenomena. With $NH_3$ on FE $In_2Se_3$ as a typical example, it is found that the adsorption energies (absolute values) on the either side are linearly increased by the tensile strain while they are decrease under the compressive strain (blue lines in Figure 2b inset). The strain modulated gas adsorption behavior can be well understood from the shifted band alignments between frontier orbitals and band-edge states. Under 5% tensile strain, the energy levels of CBM and VBM states of $In_2Se_3$ relative to the vacuum layer are shifted downwards regardless of P ↓ or P ↑ surface as shown in the inset of Figure 2b, however the HUMO and LUMO of the gas molecule are less affected. The electron transfer barrier is therefore reduced from 0.33 eV to Z-scheme for P ↓ surface. At the opposite surface, the transfer barrier is also reduced from 1.53 to 1.00 eV. Therefore, the adsorption energies at both sides can be significantly increased by the tensile strain. Under the compressive strain (-5%), the electron transfer barriers at both sides for $NH_3$ are increased, leading to decreased adsorption energies. NO adsorption preference on P ↓ are the same as $NH_3$ and the tensile strain also enhance the adsorption strength. However, as for $NO_2$, the tensile strain will induce a barrier of electron transfer on P ↑ surface, therefore the corresponding absorption energy will be decreased on P ↑ surface but increased under compressive strain. The tunable adsorption behavior not only provides the feasible approach to manipulate and control the gas adsorption or release, but also verify the correctness of our proposed mechanism.

Since the binding performance of gas molecule is quite different in $In_2Se_3$ (P ↓ ) and (P ↑ ) surfaces, one can expect to realize adsorption/release by switching the orientation of electric polarization. As illustrated in Figure 3, we propose an effective pathway via a three-step movement of the upper three Se-In-Se layers based on a previous study.[24] When $NH_3$ is physically absorbed on the top surface (P ↑ ), only an energy barrier of 0.05 eV/unit cell (UC) needs to be overcome so that the polarization direction is reversed to chemically adsorb the ammonia molecule, namely achieving the toxic gas capture by the transition from physical to chemical adsorption.[24] However, for versa vice process, from chemical to physical adsorption transition (gas release), the chemical bonding between N-In has to be broken firstly, therefore a higher energy barrier of 0.11 eV/UC needs to be overcome. Based on the fact that the out-of-plane ferroelectricity of pure $In_2Se_3$ and the FE polarization reversal has been experimentally confirmed, it is highly expected that $NH_3$ capture and release can be released by FE switch. For NO and $NO_2$, since there is no chemical bonding between molecules and the substrate, the adsorption energies are 0.2-0.4 eV/UC within the range of the weak physical adsorption, the FE reverse can well tune the adsorption strength which can achieve by only overcoming the energy barrier of 0.07 eV/UC like the pure monolayer $In_2Se_3$.[24]

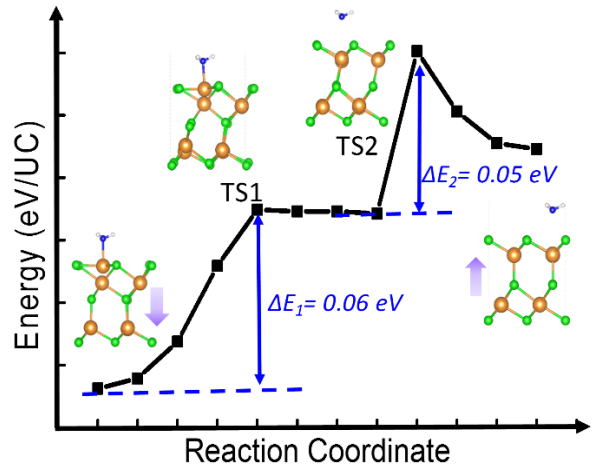

**Figure. 3** Energy profile of the effective pathway to release the $NH_3$ by reversing the electric polarization orientation of $In_2Se_3$.

So far, we have limited the adsorptions to monolayer $In_2Se_3$. To get more insight into the role of electric polarization direction in gas adsorption, we analyze the adsorptions in bilayer $In_2Se_3$. Four kinds of combinations (In$_2$Se$_3$ P ↑ ↑; P ↑ ↓; P ↓ ↑; P ↓ ↓) have been taken into considerations. Figure S3 depicts the most stable configurations of NH$_3$, NO and NO$_2$ adsorptions on all possible stacked bilayer In$_2$Se$_3$ and corresponding electronic structures (Figure S4-6) to further check the effects of the electric polarization. Taking NH$_3$ for example, compared to the scenario of NH$_3$ adsorption on In$_2$Se$_3$ P ↑, we can see that NH$_3$ is also oriented toward the In$_2$Se$_3$ bilayer with N atom pointing to the In atoms and all N-H bonds away from the surface. However, the E$_{ads}$ is slightly increased to -0.203 eV in P ↑ ↑ In$_2$Se$_3$. In contrast, in P ↑ ↓ stacked bilayer In$_2$Se$_3$, the E$_{ads}$ of NH$_3$ is decreased to -0.178 eV due to the polarization cancellation. Accordingly, the distance between N atom to the outmost Se layer is increased from 2.587 to 2.663 Å (Figure S3b).

For the adsorption of NO and NO$_2$ on bilayer In$_2$Se$_3$, the adsorption on P ↑ ↑ stacked surface is stronger than that on P ↑ ↓ stacking, which is in agreement with the tendency in NH$_3$ adsorption. In addition, the adsorption on P ↓ ↓ is more stable than that on P ↓ ↑ for the similar reason (overall cancelled polarization). These results demonstrate that under conditions of the same upper In$_2$Se$_3$ layer, the electric polarization accumulation (same polarization directions) can improve the gas adsorption compared to the opposite polarization directions.

*Magnetization switch and molecule multiferroics.* Similar to MoS$_2$ monolayer, the paramagnetic molecules NO and NO$_2$ adsorption can lead to spin polarization of electronic properties and magnetic moment. We show the spin-polarized projected density of states (PDOS) of NO adsorption as the example on surface of In$_2$Se$_3$ (P ↓) and In$_2$Se$_3$ (P ↑) in Figure 4. Obviously, for NO on In$_2$Se$_3$ (P ↓), in-gap impurity states appear near the Fermi level, leading to a total magnetic moment of 0.88 μ$_B$. Similarly, the adsorption of NO on In$_2$Se$_3$ (P ↑) also gives rise to a magnetic moment, but with a value of 1.00 μ$_B$, which means the magnetic moment of NO is sensitive to the surface conditions. It can be seen from Figure 4 that the PDOS of NO on P ↓ or P ↑ surface near the Fermi level is almost the same, however the P ↓ $In_2Se_3$ substrate has been partly spin polarized at the Fermi level, which is however absent for P ↑ surface adsorption. Similar phenomena can be observed in the case of $NO_2$ adsorption, the magnetic moment and spin polarization can be well regulated by the polarization surface (see Figure S7 and Table S1). Thus, the controllable magnetic responses on different side render $In_2Se_3$ monolayer a promising candidate to control the magnetization via the reversible FE switch, which is highly desirable in spintronics applications like magnetic switch.

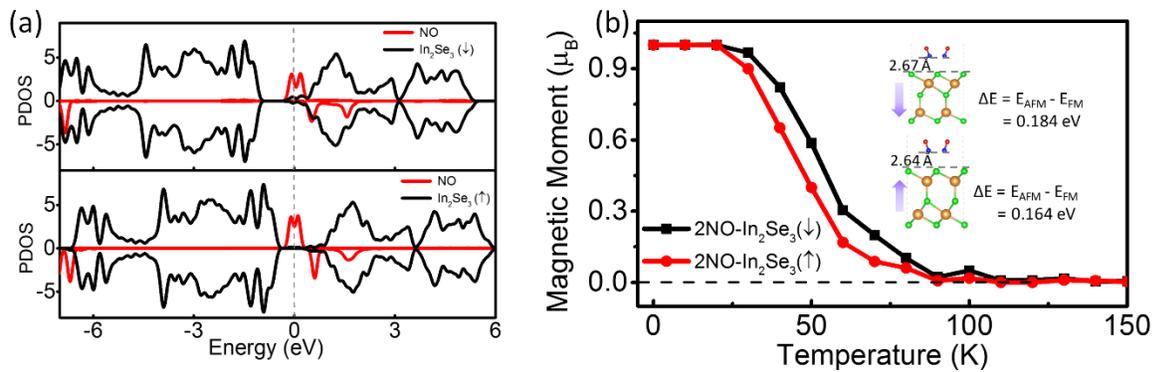

**Figure. 4** (a) Upper panel: PDOS for the NO adsorbed $In_2Se_3$ (P ↓). Lower panel: PDOS for the NO adsorbed $In_2Se_3$ (P ↑). (b) Variation of the total magnetic moment of NO- $In_2Se_3$ (P ↓) system (black line) and NO- $In_2Se_3$ (P ↑) system (red line) with respect to the temperature. Insets are side views of two NO molecules adsorbed $In_2Se_3$ (P ↓) and $In_2Se_3$ (P ↑). Energy difference (ΔE) between AFM and FM order are also shown.

To explore the magnetic coupling at higher concentration, we increased magnetic molecule coverage. When one additional $NO_2$ is placed on the surface regardless the polarization, they will expel each other, rendering the magnetic coupling between two $NO_2$ molecules rather weak and energy degeneracy of antiferromagnetic (AFM) and ferromagnetic (FM) states (see Figure S8). In contrast, the additional NO molecule prefers to stay close to neighboring NO with 2.02 Å (after relaxation). More importantly, the FM coupling between these two NO molecules is much more preferred compared with the AFM coupling regardless the adsorption surfaces. It is interesting to note that the polarization direction has also significant effects on the magnetic coupling strength; when both NO molecules are adsorbed on the top surface (P ↓), the energy difference of FM and AFM is 0.184 eV/supercell; the value becomes 0.164 eV/supercell when they are adsorbed on the opposite side (P ↑). Although the FM coupling is always the magnetic ground state regardless of the FE surface, the different magnetic coupling strength will significantly affect the Curie temperatures (Tc). To study the spin dynamics, Monte Carlo simulations were performed with the Wolff algorithm. The spin Hamiltonian based on the 2D Ising model is considered as $\hat{H} = -\sum_{i,j} J \hat{m}_i \cdot \hat{m}_j$, where $J$ is the nearest neighbor exchange parameter, $m_i$ and $m_j$ denote the magnetic moments at site $i$ and $j$. We note that the $T_C$ of NO on bottle surface (P ↑) is 50 K which is close to the liquid nitrogen temperature (77 K). However, when NO molecules on the opposite surface, Tc is increased obviously compared with that on the P ↑ surface, owing to the larger energy difference between the ferromagnetic and antiferromagnetic states. The polarization can not only affect the adsorption behavior, magnetic moments, but also has the significant impact on the Curie temperate. Our calculated Tc for NO molecules adsorbed on ferroelectric $In_2Se_3$ layer is comparable with or slightly higher than the recently reported values for 2D $CrI_3$ (~45 K)[34] and $Cr_2Ge_2Te_6$ (~25 K),[35] which indicates the feasibility to be observed and demonstrated from experiments.

With the intrinsic ferroelectricity of $In_2Se_3$ monolayer, we can see that the NO molecule adsorption on the surface will lead to the coexistence of ferroelectricity and ferromagnetism, namely the phenomena of 2D multiferroics. Although the origins of these two ferroic states are different, it belongs to the type I multiferroics, the electromagnetic coupling is not very weak since we can see the controllability of magnetic moment and Tc via FE polarization switch. With the NO coverage further increases, the ferromagnetic coupling and different magnetic behaviors on two FE surfaces can be well preserved as seen from Figures S9, implying the strong robustness of our findings. The multiferroics which combines 2D ferroelectricity and molecule ferromagnetism and the unique tunability by FE switch render the system being of great importance in next generation electronics design like non-volatile memory storages.[36-39]

**Conclusion.** In summary, based on first principles calculations, we found that FE monolayer $In_2Se_3$ can act as a reversible gas sensing substrate. The adsorption energies and electron transfers of $NH_3$, NO and $NO_2$ are significantly different on two FE surfaces, which may offer a feasible approach to capture or release the gases by FE switch after overcoming a small energy barrier. The synergistic effects from potential differences and relative band alignments between the frontier molecular orbitals of gas and band-edge states of the substrate are responsible for the observed phenomena, which provides a universal guideline for understanding of gas adsorption behavior and strain effects. More interestingly, paramagnetic molecules NO and $NO_2$ adsorption on the surface can lead to the appearance of magnetism, the magnetic moment is also tunable by ferroelectric switch. At higher coverage, NO molecules prefer the ferromagnetic coupling, the estimated Tc is around 50K, leading to the coexistence of ferroelectricity and ferromagnetism (2D multiferroics). The reversible gas sensing and the revealed molecule multiferroics render ferroelectric $In_2Se_3$ layer promising in toxic gas removal and spintronics applications.

**Computational Methods.** The density functional theory (DFT) calculations were performed by using the Vienna ab initio simulation package (VASP).[40, 41] The exchange−correlation functional was treated with the generalized gradient approximation (GGA) in the Perdew–Burke–Ernzerhof (PBE) form.[42, 43] The projected augmented wave (PAW) method was used to represent the electron-ion interactions. The cut-off energy was set to 500 eV, and a vacuum space greater than 20 Å was applied perpendicular to the sheet to avoid the interaction between neighboring layers. The atomic positions in all structures were relaxed until the convergence criteria of force and energy are less than 0.01 eV/Å and $10^{-5}$ eV, respectively. A $\Gamma$-centered Monkhorst-Pack $k$-point grids[44] of $5 \times 5 \times 1$ for geometry optimization and $9 \times 9 \times 1$ for electronic properties investigation. DFT-D3 method[45] was used to incorporate the long range van der Waals interaction, while the climbing-image nudged elastic band (CI-NEB) method was adopted to compute the energy barrier.[46] The adsorption energy ($E_{ads}$) of the gas molecule on the $In_2Se_3$ sheet was identified as $E_{ads} = E_{total}$-$E_{gas}$-$E_{In2Se3}$, where $E_{total}$, $E_{gas}$ and $E_{In2Se3}$ are energies of $In_2Se_3$ sheet with adsorbed gas molecule, isolated gas molecule and free $In_2Se_3$, respectively.

ASSOCIATED CONTENT

The following files are available free of charge.

Electrostatic potential; energy level; charge density difference, adsorption energies, electronic properties, magnetic moment of gas adsorbed bilayer $In_2Se_3$; PDOS for $NO_2$ adsorbed $In_2Se_3$; geometrics and energies of FM, AFM ,NM states of NO and $NO_2$ at higher concentration. (PDF)

AUTHOR INFORMATION

**Corresponding Author**

*E-mail: liangzhi.kou@qut.edu.au

**ORCID**

Xiao Tang: 0000-0003-3618-3817

Aijun Du: 0000-0002-3369-3283

Liangzhi Kou: 0000-0002-3978-117X

**Author Contributions**

The manuscript was written through contributions of all authors. All authors have given approval to the final version of the manuscript.

**Notes**

The authors declare no competing financial interest.

ACKNOWLEDGMENT

We acknowledge the grants of high-performance computer time from computing facility at the Queensland University of Technology, the Pawsey Supercomputing Centre and Australian National Facility. L.K. gratefully acknowledges financial support by the ARC Discovery Project (DP190101607).

TOC

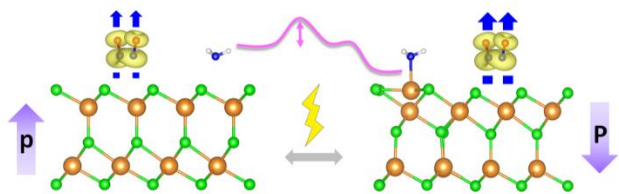